\documentclass[letterpaper,twocolumn,showpacs,pra,aps,floatfix]{revtex4}
\usepackage{graphicx}
\usepackage{array}
\usepackage{amsthm}
\usepackage{amsmath}
\usepackage{amssymb}
\usepackage{subfig}
\usepackage{calc}

\begin{document}

\title{General monogamy of Tsallis $q$-entropy entanglement in multiqubit systems}
\author{Yu Luo}
%\email{luoyu@snnu.edu.cn}
\author{Tian Tian}
%\email{luoyu@snnu.edu.cn}
\author{Lian-He Shao}
%\email{snnulhs@gmail.com}
\author{Yongming Li}
%\email{liyongm@snnu.edu.cn}

\affiliation{College of Computer Science, Shaanxi Normal University, Xi'an, 710062, China}
\date{\today}

\begin{abstract}
In this paper, we study the monogamy inequality of Tsallis $q$-entropy entanglement. We first provide an analytic formula of Tsallis $q$-entropy entanglement in two-qubit systems for $\frac{5-\sqrt{13}}{2}\leq q\leq\frac{5+\sqrt{13}}{2}.$ The analytic formula of Tsallis $q$-entropy entanglement in $2\otimes d$ system is also obtained and we show that Tsallis $q$-entropy entanglement satisfies a set of hierarchical monogamy equalities. Furthermore, we prove the squared Tsallis $q$-entropy entanglement follows a general inequality in the qubit systems.  Based on the monogamy relations, a set of multipartite entanglement indicators is constructed, which can detect all genuine multiqubit entangled states even in the case of $N$-tangle vanishes. Moreover, we study some examples in multipartite higher-dimensional system for the monogamy inequalities.
\end{abstract}
\eid{identifier}
\maketitle

\section{Introduction}
Multipartite entanglement is an important physical resource in quantum mechanics, which can be used in quantum computation, quantum communication and quantum cryptography. One of the most surprising phenomena for multipartite entanglement is that the monogamy property, which quantifies the resources of quantum entanglement can not be shared freely between different constituents in a multipartite system. Monogamy property may be as fundamental as the no-cloning theorem~\cite{Bruss99,Coffman00,Osborne06,Kay09}. A simple example of monogamy property can be interpreted as the amount of entanglement between $A$ and $B$, plus the amount of entanglement between $A$ and $C$, cannot be greater than the amount of entanglement between $A$ and the pair $B$$C$. Monogamy property has been considered in many areas of physics: One can estimate the quantity of information captured by an eavesdropper about the secret key to be extracted in quantum cryptography~\cite{Osborne06,Barrett05}, the frustration effects observed in condensed matter physics~\cite{Ma11,Amico08}, and even in black-hole physics~\cite{Susskind13,Lloyd14}.

Monogamy property of various entanglement measures have been discovered. Coffman $et$ $al.$ first considered three qubits $A$,$B$ and $C$ which may be entangled with each other~\cite{Coffman00}, who showed that the squared concurrence $\mathcal{C}^2$ follows this monogamy inequality. Osborne $et$ $al.$ proved the squared concurrence follows a general monogamy inequality for the $N$-qubit system~\cite{Osborne06}. Different kinds of monogamy inequalities for concurrence have been noted in Refs.~\cite{Ou08,LiMing16,Zhu14,Zhu15,Eltschka15}. Some similar monogamy inequalities were also discussed for entanglement of formation~\cite{Zhu14,Bai1401,Bai1408}, negativity~\cite{Kim09,Ou0700,Luo15,He15,Tian16}, relative entropy entanglement~\cite{Li14,Lancien16}, continuous variable systems~\cite{Hiroshima07,Adesso07,Adesso08}, Renyi $\alpha$-entropy entanglement~\cite{Kim100,Song16}, and Tsallis $q$-entropy entanglement~\cite{Kim10,Kim1603}. The monogamy property of other physical resources has also been discussed, such as discord~\cite{Bai13,Streltsov12}, and steering~\cite{QYHe13,Pramanik14}.

Tsallis $q$ entropy is an important entropic measure, which can be used in many areas of quantum information theory~\cite{Rajagopal05,Rossignoli02,Batle02,Abe01,Tsallis01,Barranco99}. In this paper, we study the monogamy inequality of Tsallis $q$-entropy entanglement (TEE). We first provide an analytic formula of TEE in two-qubit systems for $\frac{5-\sqrt{13}}{2}\leq q\leq\frac{5+\sqrt{13}}{2}.$ The analytic formula of TEE in the $2\otimes d$ system is also obtained and we show that TEE satisfies a set of hierarchical monogamy equalities. Furthermore, we prove the squared TEE follows a general inequality in the qubit systems. As a corollary, we provide that the $\alpha$th power of TEE satisfies the monogamy inequality for $\alpha\geq2$.  Based on the monogamy relations, a set of multipartite entanglement indicators is constructed, which can detect all genuine multiqubit entangled states even in the case of $N$-tangle vanishes. Moreover, we study some examples in the multipartite higher-dimensional system for the monogamy inequalities.

This paper is organized as follows. In Sec.~\ref{sec:TEE}, we recall the definition of TEE and entanglement of formation. In Sec.~\ref{sec:main result}, we discuss the monogamy properties of TEE. In Sec.~\ref{sec:indicator}, we construct a set of multipartite entanglement indicators, and analysis of some examples. In Sec.~\ref{sec:higher dim}, we study some examples in the multipartite higher-dimensional system for the monogamy inequalities. We summarize our results in Sec.~\ref{sec:conclusion}.

\section{Quantifying Entanglement by Tsallis $q$-entropy}\label{sec:TEE}
Quantifying entanglement is an important problem in quantum information. Given a bipartite state $\rho_{AB}$ in the Hilbert space $\mathcal{H_A}\otimes\mathcal{H_B}$. The Tsallis-$q$ entropy is defined as~\cite{Tsallis88}
\begin{equation}
T_q(\rho)=\frac{1}{q-1}(1-Tr\rho^q)
\end{equation}
for any $q >0$ and $q\neq 1$. When $q$ tends to 1, the Tsallis $q$-entropy $T_q(\rho)$ converges to its von Neumann entropy~\cite{Nielsen00}: $\lim_{q\to 1}T_q(\rho)=-Tr(\rho\ln\rho)$. For any pure state $|\psi_{AB}\rangle$, the TEE is defined as
\begin{equation}
\mathcal{T}_{q}(|\psi_{AB}\rangle)=T_{q}(\rho_A)
\end{equation}
for any $q >0$. For a mixed state $\rho_{AB}$, the TEE can be defined as
\begin{equation}
\mathcal{T}_{q}(\rho_{AB})=\min\sum_ip_i\mathcal{T}_{q}(|\psi_{AB}^i\rangle),
\end{equation}
for any $q>0$, where the minimum is taken over all possible pure state decompositions $\{p_i,\psi_{AB}^i\}$ of $\rho_{AB}.$ TEE can be viewed as a general entanglement of formation when $q$ tends to 1. The entanglement of formation is defined as~\cite{Bennett9601,Bennett9602}
\begin{equation}
E_f(\rho_{AB})=\min\sum_ip_iE_f(|\psi^i_{AB}\rangle),
\end{equation}
where $E_f(|\psi^i_{AB}\rangle)=-Tr\rho_A^i\ln\rho_A^i=-Tr\rho_B^i\ln\rho_B^i$ is the von Neumann entropy, the minimum is taken over all possible pure state decompositions $\{p_i,\psi_{AB}^i\}$ of $\rho_{AB}.$ In Re.~\cite{Wootters98}, Wootters derived an analytical formula for a two-qubit mixed state $\rho_{AB}$
\begin{equation}
E_f(\rho_{AB})=H(\frac{1+\sqrt{1-\mathcal{C}_{AB}^2}}{2}),
\end{equation}
where $H(x)=-x\ln x-(1-x)\ln (1-x)$ is the binary entropy and $\mathcal{C}_{AB}=\max\{0,\lambda_1-\lambda_2-\lambda_3-\lambda_4\}$ is the concurrence of $\rho_{AB}$, with $\lambda_i$ being the eigenvalues, in decreasing order, of matrix $\sqrt{\rho_{AB}(\sigma_y\otimes\sigma_y)\rho_{AB}^{*}(\sigma_y\otimes\sigma_y)}$~\cite{Wootters98}.

In particular, Kim found $\mathcal{T}_{q}(\rho_{AB})$ has an analytical formula for a two-qubit mixed state, which can be expressed as a function of the squared concurrence $\mathcal{C}_{AB}^2$ for $1\leq q \leq 4$~\cite{Kim10}
\begin{equation}\label{eq:TEE}
\mathcal{T}_{q}(\rho_{AB})=f_q(\mathcal{C}_{AB}^2),
\end{equation}
where the function $f_q(x)$ has the form
\begin{equation}\label{eq:fqx}
f_q(x)=\frac{1}{q-1}[1-(\frac{1+\sqrt{1-x}}{2})^q-(\frac{1-\sqrt{1-x}}{2})^q].
\end{equation}
In this paper, we further prove that the analytical formula also holds for $q\in[\frac{5-\sqrt{13}}{2},\frac{5+\sqrt{13}}{2}],$ where $\frac{5-\sqrt{13}}{2}\approx 0.697$ and $\frac{5+\sqrt{13}}{2}\approx 4.302.$ We refer the interested readers to Appendices A for the detailed calculation.

\section{Monogamy of TEE in multiqubit systems}\label{sec:main result}
Before presenting our main results, we have the following properties for TEE $f_{q}(\mathcal{C}^2)$.

\emph{Property 1:} The squared Tsallis $q$-entropy entanglement $f_{q}^2(\mathcal{C}^2)$ is an increase monotonic and convex function of the squared concurrence $\mathcal{C}^2$ for any two-qubit mixed states, where $q\in[\frac{5-\sqrt{13}}{2},\frac{5+\sqrt{13}}{2}].$

\emph{Property 2:} The Tsallis $q$-entropy entanglement $f_{q}(\mathcal{C}^2)$ is an increase monotonic and concave function of the squared concurrence $\mathcal{C}^2$, where $q\in[\frac{5-\sqrt{13}}{2},2]\cup[3,\frac{5+\sqrt{13}}{2}].$

We refer the interested readers to Appendixes B and C for the detailed proof for properties above. The region of $q$ we considered for the properties is $q\in[\frac{5-\sqrt{13}}{2},\frac{5+\sqrt{13}}{2}].$

It's well known that for any pure state in a $2\otimes d$ system, TEE has an analytical expression for $q>0$~\cite{Kim10}. We have the following result for any mixed state in a $2\otimes d$ system:

{\sf Theorem 1}~. For a mixed state $\rho_{A\textbf{C}}$ in a $2\otimes d$ system, TEE has an analytical expression
\begin{eqnarray}\label{eq:The2}
\mathcal{T}_{q}(\rho_{A\textbf{C}})= f_q[\mathcal{C}^2(\rho_{A\textbf{C}})],
\end{eqnarray}
for $q\in[\frac{5-\sqrt{13}}{2},2]\cup[3,\frac{5+\sqrt{13}}{2}].$

\emph{Proof.} First, we should prove $\mathcal{T}_{q}(\rho_{A\textbf{C}})\leq f_{q}[\mathcal{C}^2(\rho_{A\textbf{C}})]$. For $q\in[\frac{5-\sqrt{13}}{2},2]\cup[3,\frac{5+\sqrt{13}}{2}]$, consider a mixed state $\rho_{A\textbf{C}}$ in a $2\otimes d$ system.  We use an optimal convex decomposition $\{p_i,|\phi^i_{A\textbf{C}}\rangle \}$ for the TEE $\mathcal{T}_{q}(\rho_{A\textbf{C}})$:

\begin{eqnarray}\label{eq:The21}
\mathcal{T}_{q}(\rho_{A\textbf{C}})&=&\nonumber \sum_ip_i\mathcal{T}_{q}(|\phi^i_{A\textbf{C}}\rangle) \\\nonumber
&=&\sum_ip_if_{q}[\mathcal{C}^2(|\phi^i_{A\textbf{C}}\rangle)] \\\nonumber
&\leq&\sum_js_jf_{q}[\mathcal{C}^2(|\psi^j_{A\textbf{C}}\rangle)] \\\nonumber
&\leq& f_{q}[\sum_js_j\mathcal{C}^2(|\psi^j_{A\textbf{C}}\rangle)] \\
&=& f_{q}[\mathcal{C}^2(\rho_{A\textbf{C}})],
\end{eqnarray}
where we have used an optimal convex decomposition $\{s_j,|\psi^j_{A\textbf{C}}\rangle \}$ for concurrence $\mathcal{C}^2(\rho_{A\textbf{C}})=\min\sum_js_j\mathcal{C}^2(|\psi^j_{A\textbf{C}}\rangle)$ in the first inequality. The second inequality holds is due to the function $f_{q}(\mathcal{C}^2)$ is a concave function of the squared concurrence $\mathcal{C}^2$ for $q\in[\frac{5-\sqrt{13}}{2},2]\cup[3,\frac{5+\sqrt{13}}{2}].$

Secondly, we will prove $\mathcal{T}_{q}(\rho_{A\textbf{C}})\geq f_{q}[\mathcal{C}^2(\rho_{A\textbf{C}})]$. We can obtain

\begin{eqnarray}\label{eq:The22}
\mathcal{T}_{q}(\rho_{A\textbf{C}})&=&\nonumber \sum_ip_i\mathcal{T}_{q}(|\phi^i_{A\textbf{C}}\rangle) \\\nonumber
&=&\sum_ip_if_{q}[\mathcal{C}(|\phi^i_{A\textbf{C}}\rangle)] \\\nonumber
&\geq& f_{q}\{[\sum_js_j\mathcal{C}(|\psi^j_{A\textbf{C}}\rangle)]^2\} \\\nonumber
&\geq& f_{q}\{[\sum_kr_k\mathcal{C}(|\psi^j_{A\textbf{C}}\rangle)]^2\} \\
&=& f_{q}[\mathcal{C}^2(\rho_{A\textbf{C}})],
\end{eqnarray}
where the first inequality holds due to the convexity of $f_q(\mathcal{C}^2)$ as the function of concurrence $\mathcal{C}$ for $q>0$ (see Appendix A), and we have used the optimal convex decomposition $\{r_k,|\psi^k_{A\textbf{C}}\rangle \}$ for concurrence $\mathcal{C}(\rho_{A\textbf{C}})=\min\sum_kr_k\mathcal{C}(|\psi^k_{A\textbf{C}}\rangle)$ in the second inequality, thus proving Theorem 1. \qquad \qquad \qquad \qquad \qquad \qquad \qquad $\square$

A straightforward corollary of Theorem 1 is

{\sf Corollary 1}~. For any mixed state in a $2\otimes d$ system, TEE obeys the following relation:

\begin{eqnarray}\label{eq:Coro1}
\mathcal{T}_{q}(\rho_{A\textbf{C}})\geq f_{q}[\mathcal{C}^2(\rho_{A\textbf{C}})],
\end{eqnarray}
where $q>0$.

The Eq.~(\ref{eq:Coro1}) provides a lower bound for TEE in the $2\otimes d$ system.

Now we will study the monogamy property of TEE. We have the following theorem first:

{\sf Theorem 2}~. For a mixed state $\rho_{A|B\textbf{C}}$ in a $2\otimes2\otimes2^{N-2}$ system,
the following monogamy inequality holds:
\begin{equation}\label{eq:The3}
\mathcal{T}_{q}^2(\rho_{A|B\textbf{C}})\geq\mathcal{T}_{q}^2(\rho_{AB})+\mathcal{T}_{q}^2(\rho_{A\textbf{C}}),
\end{equation}
where $q\in[\frac{5-\sqrt{13}}{2},2]\cup[3,\frac{5+\sqrt{13}}{2}].$

\emph{Proof.} Consider a mixed state $\rho_{A|B\textbf{C}}$ in a $2\otimes2\otimes2^{N-2}$ system for $q\in[\frac{5-\sqrt{13}}{2},2]\cup[3,\frac{5+\sqrt{13}}{2}],$ from the Eq.~(\ref{eq:The2}) we have:
\begin{eqnarray}
\mathcal{T}_{q}^2(\rho_{A|B\textbf{C}})&=&\nonumber f_{q}^2[\mathcal{C}^2(\rho_{A|B\textbf{C}})] \\\nonumber
&\geq& f_{q}^2[\mathcal{C}^2(\rho_{AB})+\mathcal{C}^2(\rho_{A\textbf{C}})] \\\nonumber
&\geq& f_{q}^2[\mathcal{C}^2(\rho_{AB})]+f_{q}^2[\mathcal{C}^2(\rho_{A\textbf{C}})] \\\nonumber
&=& \mathcal{T}_{q}^2(\rho_{AB})+\mathcal{T}_{q}^2(\rho_{A\textbf{C}}),
\end{eqnarray}
where the first inequality holds is due to $f^2_q(x)$ is an increase monotonic function of the squared concurrence $\mathcal{C}^2$ and $\mathcal{C}^2(\rho_{A|B\textbf{C}})\geq\mathcal{C}^2(\rho_{AB})+\mathcal{C}^2(\rho_{A\textbf{C}})$ for concurrence~\cite{Osborne06}. The second inequality holds is due to convexity of $f_{q}^2(\mathcal{C}^2)$ as a function of $\mathcal{C}^2$. \qquad \qquad \qquad \qquad \qquad \qquad \qquad $\square$

From Theorem 2, a set of hierarchical monogamy inequalities of $\mathcal{T}_{q}^2(\rho_{A_1|A_2\ldots{A_N}})$ holds for any $N$-qubit mixed state $\rho_{A_1A_2\ldots{A_N}}$ in $k$-partite cases with $k=\{3,4,\ldots,N\}$:
\begin{equation}\label{eq:The31}
\mathcal{T}_{q}^2(\rho_{A_1|A_2\ldots{A_N}})\geq\sum_{i=2}^{k-1}\mathcal{T}_{q}^2(\rho_{A_1A_i})+\mathcal{T}_{q}^2(\rho_{A_1|A_k\ldots{A_N}}),
\end{equation}
where $q\in[\frac{5-\sqrt{13}}{2},2]\cup[3,\frac{5+\sqrt{13}}{2}].$ These set of hierarchical relations can be used to detect the multipartite entanglement in these $k$-partite. When $k=N$, we have following monogamy inequality for $q\in[\frac{5-\sqrt{13}}{2},2]\cup[3,\frac{5+\sqrt{13}}{2}]$

\begin{equation}\label{eq:Mono1}
\mathcal{T}^{2}_{q}(\rho_{A_1|A_2\ldots A_N})\geq \mathcal{T}^{2}_{q}(\rho_{A_1A_2})+\cdots+\mathcal{T}^{2}_{q}(\rho_{A_1A_N}).
\end{equation}

One can wonder whether the monogamy inequality Eq.~(\ref{eq:Mono1}) still holds for $q\in[2,3]$. Here, we give an affirmative answer. In Ref.~\cite{Kim10}, the author proved the following inequality for $q\in[2,3]$

\begin{equation}\label{eq:Mono2}
\mathcal{T}_{q}(\rho_{A_1|A_2\ldots A_N})\geq \mathcal{T}_{q}(\rho_{A_1A_2})+\cdots+\mathcal{T}_{q}(\rho_{A_1A_N}),
\end{equation}
which is easy to check that the inequality Eq.~(\ref{eq:Mono1}) also holds for $q\in[2,3]$ from Eq.~(\ref{eq:Mono2}). Thus we have following result.

{\sf Theorem 3}~. For a mixed state $\rho_{A_1A_2\ldots A_N}$ in an $N$-qubit system, the following monogamy inequality holds
\begin{equation}\label{eq:The1}
\mathcal{T}^{2}_{q}(\rho_{A_1|A_2\ldots A_N})\geq \mathcal{T}^{2}_{q}(\rho_{A_1A_2})+\cdots+\mathcal{T}^{2}_{q}(\rho_{A_1A_N}),
\end{equation}
for $q\in[\frac{5-\sqrt{13}}{2},\frac{5+\sqrt{13}}{2}].$

Bai $et$ $al.$ show that the squared entanglement of formation follows the general monogamy inequality in multiqubit systems~\cite{Bai1401,Bai1408}. Here, we prove the monogamous property of multiqubit entanglement can also be characterized in terms of squared TEE, where the monogamy inequality in terms of the squared entanglement of formation can be viewed as a special case for $q=1$.

As a result of Theorem 3, we also have the following corollary:

{\sf Corollary 2}~. For a mixed state $\rho_{A_1A_2\ldots A_N}$ in an $N$-qubit system, the $\alpha$th power of TEE satisfies the monogamy inequality
\begin{equation}\label{eq:alpha}
\mathcal{T}^{\alpha}_{q}(\rho_{A_1|A_2\ldots A_N})\geq \mathcal{T}^{\alpha}_{q}(\rho_{A_1A_2})+\cdots+\mathcal{T}^{\alpha}_{q}(\rho_{A_1A_N}),
\end{equation}
for $\alpha\geq2$ and $q\in[\frac{5-\sqrt{13}}{2},\frac{5+\sqrt{13}}{2}]$.

The proof can be found in Appendices D. We can view the coefficient $\alpha$ as a kind of assigned weight to regulate the monogamy property~\cite{Luo15,Regula14,Salini14}.

\section{A New Kind of Multipartite Entanglement Indicator}\label{sec:indicator}
Based on the Eq.~(\ref{eq:The1}), we can construct a class of multipartite entanglement indicator for $q\in[\frac{5-\sqrt{13}}{2},\frac{5+\sqrt{13}}{2}]$

\begin{equation}\label{eq:Indic1}
\tau_q(\rho_{A_1|A_2\ldots{A_N}})=\min\sum_ip_i\tau_q(|\psi_{A_1|A_2\ldots{A_N}}^i\rangle),
\end{equation}
where the minimum is taken over all possible pure state decompositions $\{p_i,\psi_{A_1|A_2\ldots{A_N}}^i\}$ of $\rho_{A_1A_2\ldots{A_N}}$ and $\tau_q(|\psi_{A_1|A_2\ldots{A_N}}^i\rangle=\mathcal{T}^{2}_{q}(\psi_{A_1|A_2\ldots A_N}^i)-\sum_{j=2}^N \mathcal{T}^{2}_{q}(\rho_{A_1A_j}^i)$. Use the concavity of Tsallis $q$-entropy for $q>0$~\cite{Raggio95}, and follow the method of deriving the squared entanglement of formation in Re.~\cite{Bai1401}, we have following result:

{\sf Theorem 4}~. For any three-qubit mixed state $\rho_{ABC}$, the multipartite entanglement indicator $\tau_q(\rho_{A|BC})$ is zero if and only if $\rho_{ABC}$ is biseparable, i.e., $\rho_{ABC}=\sum_ip_i\rho_{AB}^i\otimes\rho_{C}^i+\sum_jp_j\rho_{AC}^j\otimes\rho_{B}^j+\sum_kp_k\rho_{A}^k\otimes\rho_{BC}^k.$

We will show some examples as blow.

{\sf Example 1}. Coffman $et$ $al$ considered a three-qubit general W state $|W\rangle_G=\sin\theta\cos\phi|001\rangle+\sin\theta\sin\phi|010\rangle+\cos\phi|100\rangle$ where $0\leq\theta\leq\pi$ and $0\leq\phi\leq2\pi$, they found the three tangle vanishes for every parameter $\theta$ and $\phi$~\cite{Coffman00}. In this case, we consider the multipartite entanglement indicator shown in Eq.~(\ref{eq:Indic1}). For this state, the value of $\tau_q(|W\rangle_G)$ can be given by its analytical formula Eq.~(\ref{eq:TEE}). In Figs. 1-4, we plot the indicator $\tau_q(|W\rangle_G)$ for $q=0.7,1,2.5,4.3$. The indicator $\tau_q(|W\rangle_G)$ shows that the $\tau_q(|W\rangle_G)$ is nonnegative for $0\leq\theta\leq\pi$ and $0\leq\phi\leq2\pi$, which vanishes when $|W\rangle_G$ is separable, thus the situation of
 $\theta=\frac{\pi}{2},\pi$ and $\phi=\frac{\pi}{2},\pi,\frac{3\pi}{2},2\pi$. For example, when $\theta=\frac{\pi}{2}$, the related state becomes $|W\rangle_G=\cos\phi|001\rangle+\sin\phi|010\rangle$ which is separable.

\begin{figure}
\begin{center}
\begin{minipage}[c]{0.25\textwidth}
\centering
\includegraphics[angle=0,width=3cm,height=3cm]{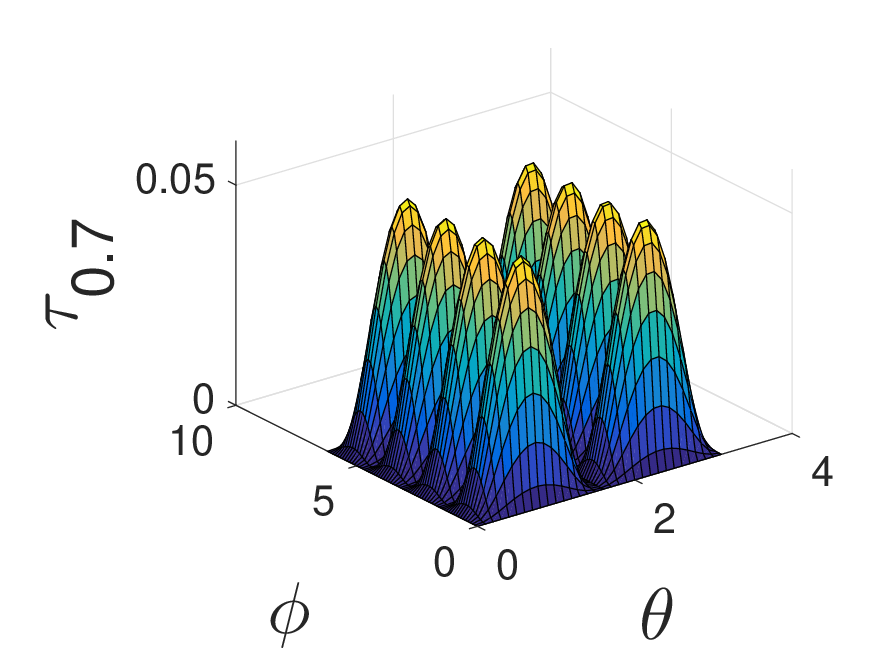}
\centering
\caption{(color online) The indicator $\tau_{0.7}(|W\rangle_G)$.}
\label{fig:levfig}
\end{minipage}%
\begin{minipage}[c]{0.25\textwidth}  \centering\includegraphics[angle=0,width=3cm,height=3cm]{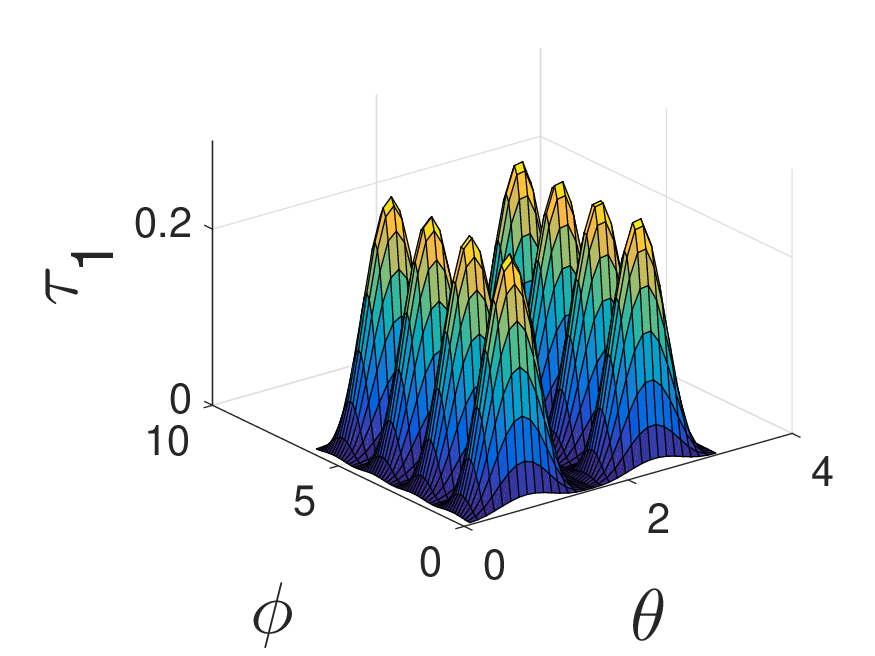}
\caption{(color online) The indicator $\tau_{1}(|W\rangle_G)$. }
\label{fig:levfig}
\end{minipage}
\end{center}
\end{figure}

\begin{figure}
\begin{center}
\begin{minipage}[c]{0.25\textwidth}
\centering
\includegraphics[angle=0,width=3cm,height=3cm]{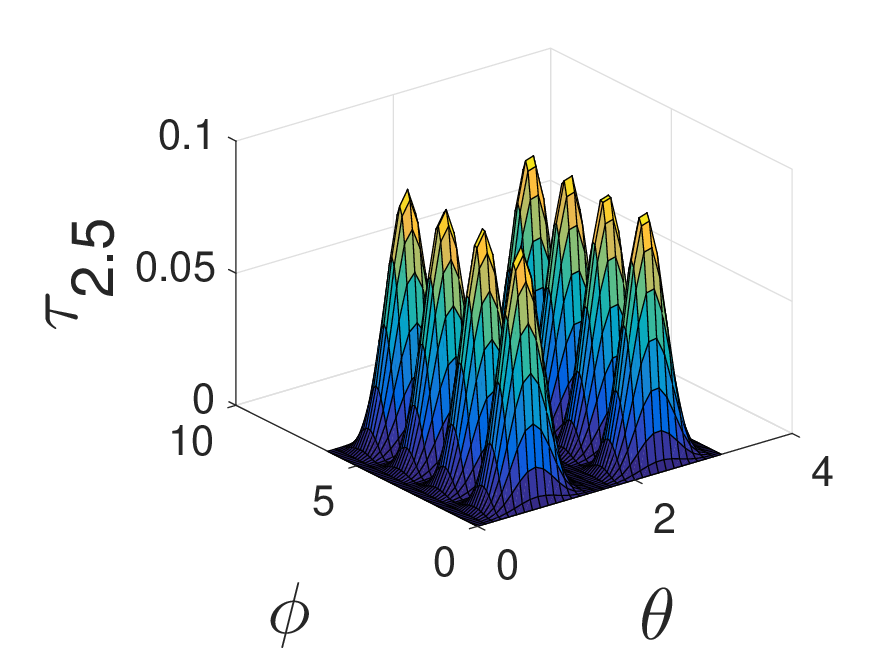}
\centering
\caption{(color online) The indicator $\tau_{2.5}(|W\rangle_G)$.}
\label{fig:levfig}
\end{minipage}%
\begin{minipage}[c]{0.25\textwidth}  \centering\includegraphics[angle=0,width=3cm,height=3cm]{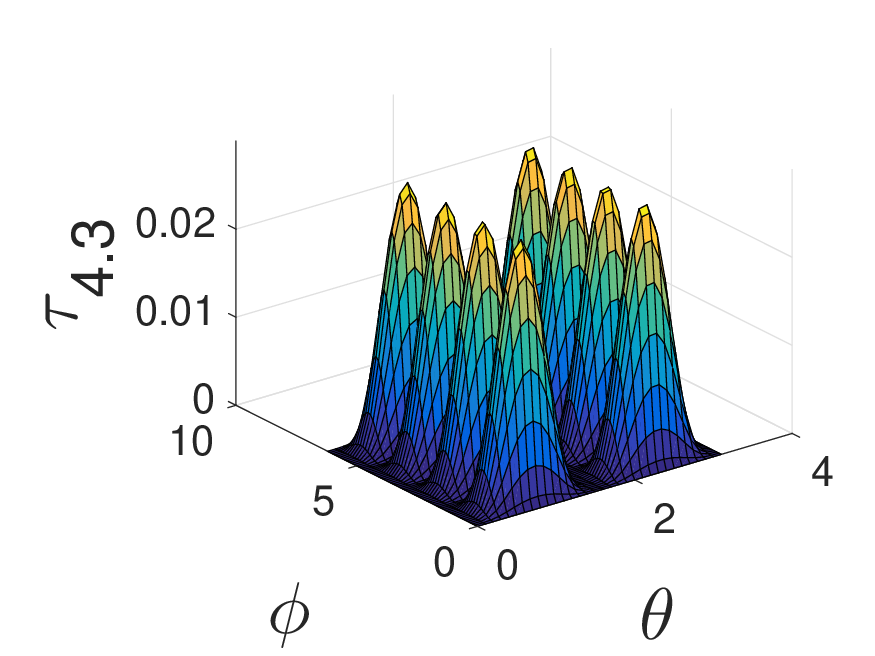}
\caption{(color online) The indicator $\tau_{4.3}(|W\rangle_G)$. }
\label{fig:levfig}
\end{minipage}
\end{center}
\end{figure}

{\sf Example 2}. We consider the $N$-qubit W state $|W\rangle_N=\frac{1}{\sqrt N}(|10\cdots0\rangle+|01\cdots0\rangle+|0\cdots01\rangle)$, the three-tangle can not detect the entanglement of this state. By using the multipartite entanglement indicator shown in Eq.~(\ref{eq:Indic1}), we have $\tau_q(|W\rangle_N)=f_q^2(\frac{4(N-1)}{N^2})-(N-1)f_q^2(\frac{4}{N^2})$. In Fig. 5, we plot the indicator $\tau_q(|W\rangle_N)$ for $N=3,6,9,11$ respectively. It shows that the indicator $\tau_q(|W\rangle)$ is always positive for $q\in[\frac{5-\sqrt{13}}{2},\frac{5+\sqrt{13}}{2}]$.

\begin{figure}[htbp]
\begin{center}
\includegraphics[width=6cm,height=6cm]{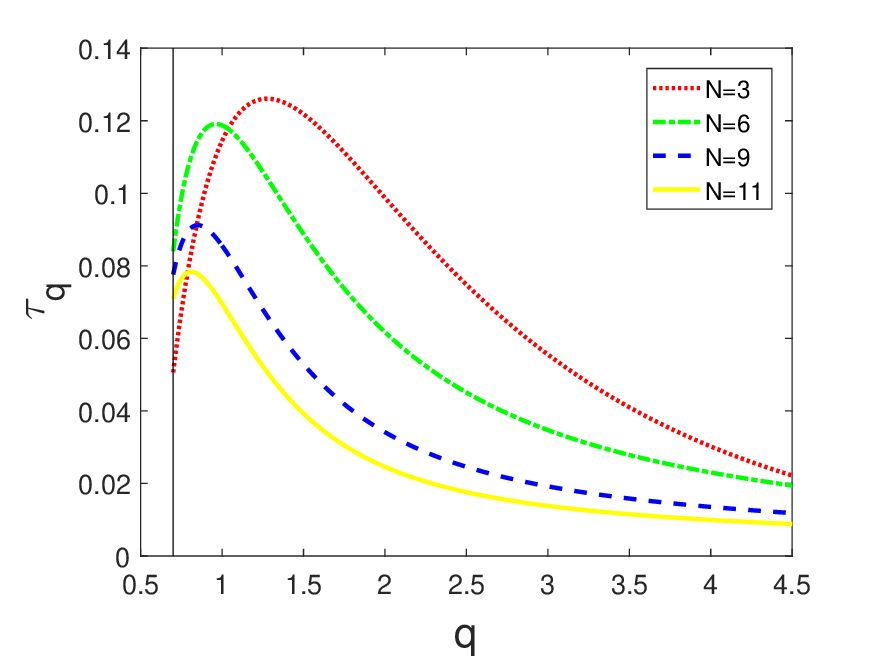}
\caption{(color online) The indicator $\tau_q(|W\rangle_N)$ is always positive for $q\in[\frac{5-\sqrt{13}}{2},\frac{5+\sqrt{13}}{2}]$.}
\end{center}
\end{figure}

\section{Monogamous examples in the multipartite higher-dimensional system}\label{sec:higher dim}
In this section, let's consider several higher-dimensional examples to illustrate the monogamy inequality of TEE in Eq.~(\ref{eq:The1}).  We define the "residual tangle" of TEE as
\begin{equation}
\tau_q(|\psi_{A_1A_2\ldots A_N}\rangle)=\mathcal{T}_{q}^{2}(\rho_{A_1|A_2\ldots A_N})-\sum_{i=2}^{N}\mathcal{T}_{q}^{2}(\rho_{A_1A_i}).
\end{equation}

{\sf Example 3}~(Bai $et$ $al.$~\cite{Bai1408}). Consider a tripartite pure state in a $4\otimes2\otimes2$ system
\begin{equation}
|\psi_{ABC}\rangle=\frac{1}{\sqrt2}(\alpha|000\rangle+\beta|110\rangle+\alpha|201\rangle+\beta|311\rangle),
\end{equation}
where $\alpha=\cos\theta$ and $\beta=\sin\theta$. Bai $et$ $al.$ point out the three-tangle is nonpositive for this state~\cite{Bai1408}. But the monogamy relation of squared TEE still works for this state when $q\in[1,\frac{5+\sqrt{13}}{2}]$:
\begin{eqnarray}
\tau_q(|\psi_{A|BC}\rangle)&=&\nonumber \mathcal{T}_{q}^2(|\psi_{A|BC}\rangle)-\mathcal{T}_{q}^2(\rho_{AB})+\mathcal{T}_{q}^2(\rho_{AC}) \nonumber\\
&=&\frac{(1-a)(1-b)}{(q-1)^2}[(1+a)(1+b)-2]\nonumber\\
&\geq&0
\end{eqnarray}
where $a=(\frac{1}{2})^{q-1}$ and $b=\alpha^{2q}+\beta^{2q}$. When $q=1$, the TEE converges to entanglement of formation, which has been discussed in Re.~\cite{Bai1408}.

{\sf Example 4}~(Ou~\cite{Ou07}). Let $|\psi_{ABC}\rangle$ be a totally antisymmetric pure state on a three-qutrit system
\begin{equation}
|\psi_{ABC}\rangle=\frac{1}{\sqrt 6}(|123\rangle-|132\rangle+|231\rangle-|213\rangle+|312\rangle-|321\rangle).
\end{equation}
Ou point out the $CKW$ inequality in Ref.~\cite{Coffman00} does not work for this state~\cite{Ou07}. However, for the squared TEE of this state
\begin{eqnarray}
\tau_q(|\psi_{A|BC}\rangle)&=&\nonumber \mathcal{T}_{q}^2(|\psi_{A|BC}\rangle)-\mathcal{T}_{q}^2(\rho_{AB})+\mathcal{T}_{q}^2(\rho_{AC}) \nonumber\\
&=&\frac{1}{(q-1)^2}[(1-(\frac{1}{3})^{q-1})^2-2(1-(\frac{1}{2})^{q-1})^2], \nonumber
\end{eqnarray}
and the TEE can still work for this state when $q\in[\frac{5-\sqrt{13}}{2},q_1]$, where $q_1\approx1.619.$

{\sf Example 5}~(Kim $et$ $al.$~\cite{Kim09}). For a pure state $|\psi_{ABC}\rangle$ in a $3\otimes2\otimes2$ system
\begin{equation}
|\psi_{ABC}\rangle=\frac{1}{6}(\sqrt2|121\rangle+\sqrt2|212\rangle+|311\rangle+|322\rangle).
\end{equation}
Kim $et$ $al$ shows that the $CKW$ inequality does not work for this state~\cite{Kim09}.

The reduced state of subsystem $A$ is
\begin{equation}
 \rho_{A}= \frac{1}{3}
\begin{pmatrix} 1 & 0 & 0  \\ 0 & 1 & 0
\\ 0 & 0 & 1  \end{pmatrix},
\end{equation}
the TEE of $\rho_{A}$ is $\mathcal{T}_{q}(|\psi_{A|BC}\rangle)=\frac{1}{q-1}[1-(\frac{1}{3})^{q-1}]$. the bipartite reduced state of subsystem $AB$ can be written as
\begin{equation}
\rho_{AB}=\frac{1}{2}(|x\rangle_{AB}\langle x|+|y\rangle_{AB}\langle y|),
\end{equation}
where
\begin{equation}
|x\rangle_{AB}=\frac{\sqrt2}{\sqrt3}|12\rangle+\frac{1}{\sqrt3}|31\rangle,
\end{equation}

\begin{equation}
|y\rangle_{AB}=\frac{\sqrt2}{\sqrt3}|21\rangle+\frac{1}{\sqrt3}|32\rangle.
\end{equation}
It can be shown that for arbitrary pure states $|\phi_{AB}\rangle=c_x|x\rangle_{AB}+c_y|y\rangle_{AB}$ with $|c_x|^2+|c_y|^2=1$, their reduced state $\rho_A=Tr_B(|\phi\rangle_{AB}\langle\phi|)$ has the same spectrum $\{0,1/3,2/3\}$. Then, the TEE of $|\phi_{AB}\rangle$ is $\mathcal{T}_{q}(|\phi_{AB}\rangle)=\frac{1}{q-1}[1-(1+2^q)(\frac{1}{3})^{q-1}].$ Thus, the TEE of $\rho_{AB}$ is $\mathcal{T}_{q}(\rho_{AB})=\frac{1}{q-1}[1-(1+2^q)(\frac{1}{3})^{q-1}].$ In the same way, the TEE of $\rho_{AC}$ is $\mathcal{T}_{q}(\rho_{AC})=\frac{1}{q-1}[1-(1+2^q)(\frac{1}{3})^{q-1}].$ We find the monogamy inequality of TEE still holds for $q\in[\frac{5-\sqrt{13}}{2},q_2]$, where $q_2\approx2.471.$

As shown in Fig. 6, we have plotted "residual tangle" $\tau_q(|\psi_{A|BC}\rangle)$ as the function of $q$ for the states of Examples 4 and 5, respectively. In the multipartite higher-dimensional system, the monogamy inequality Eq.~(\ref{eq:The1}) still works for the suitable parameter $q$.

\begin{figure}[htbp]
\begin{center}
\includegraphics[width=6cm,height=6cm]{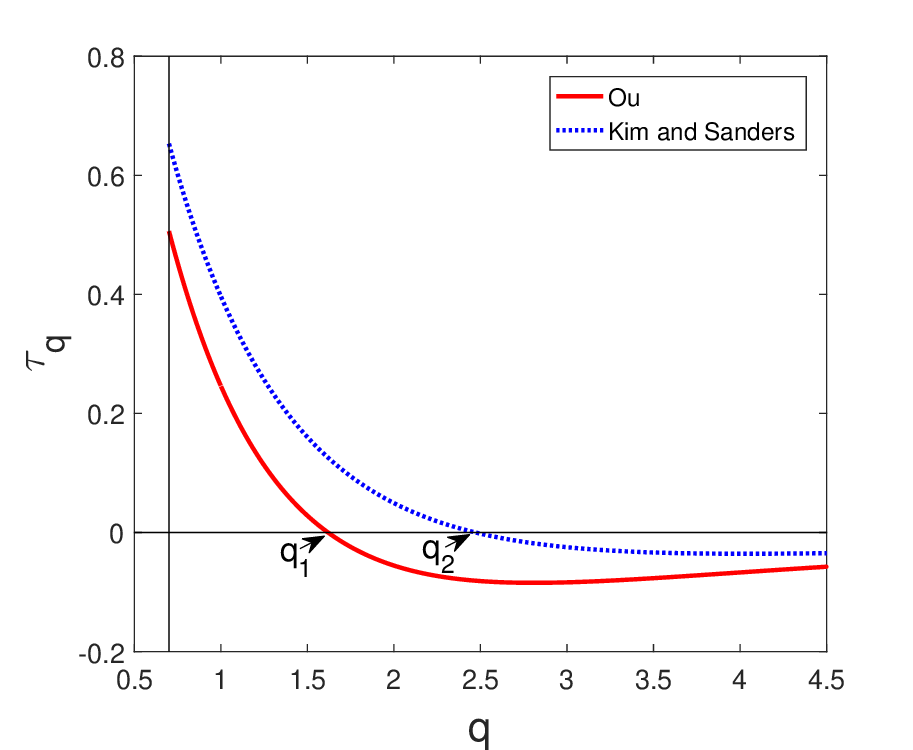}
\caption{(color online) the "residual tangle" $\tau_q(|\psi_{A|BC}\rangle)$ still works for $\frac{5-\sqrt{13}}{2}\leq q\leq q_1\approx1.619$ of Example 4 (solid red line) and for $\frac{5-\sqrt{13}}{2}\leq q\leq q_2\approx2.471$ of Example 5 (dashed blue line).
 }
\end{center}
\end{figure}

\section{conclusion}\label{sec:conclusion}

In this paper, we study the monogamy inequality of TEE. We provide an analytic formula of TEE in two-qubit systems for $\frac{5-\sqrt{13}}{2}\leq q\leq\frac{5+\sqrt{13}}{2}.$ The analytic formula of TEE in $2\otimes d$ system is also obtained and we show that TEE satisfies a set of hierarchical monogamy equalities. Furthermore, we prove the squared TEE follows a general inequality in the qubit systems. As a corollary, we provide the $\alpha$th power of TEE satisfies the monogamy inequality for $\alpha\geq2$.  Based on the monogamy relations, a set of multipartite entanglement indicators is constructed, which can detect all genuine multiqubit entangled states even in the case of $N$-tangle vanishes. Moreover, we study some examples in multipartite higher-dimensional system for the monogamy inequalities. Computing a variety of entanglement measures is $NP$-hard~\cite{Huang13}, which implies (in a rigorous sense) that the analytical formulas of TEE for general mixed states are impossible unless $P=NP$. Thus, to find a useful method to compute general entanglement measures is still a problem. We may find other methods to derive new monogamy inequalities.

For entanglement of formation, its $\alpha$-th power satisfies the monogamy inequality in Eq.~(\ref{eq:alpha}) for $\alpha\geq\sqrt2$~\cite{Zhu14}. However, the monogamy inequality of the $\alpha$-th power of TEE does not work for $\alpha\geq\sqrt2$. To see this, we can consider the three-qubit W state $|W_{A|BC}\rangle=\frac{1}{\sqrt3}(|001\rangle+|010\rangle+|100\rangle)$. Let $q=0.7$ and $\alpha=\sqrt2$, we find that $\mathcal{T}_{q}^{\alpha}(|W_{A|BC}\rangle)-\mathcal{T}_{q}^{\alpha}(\rho_{AB})-\mathcal{T}_{q}^{\alpha}(\rho_{Ac})\approx-0.087<0$. Finally, we believe our results can be used in the quantum physics.

Recently, we noted a similar work in Re.~\cite{Yuan16}.

\section{acknowledgments}
We thank Yichen Huang for sharing his paper~\cite{Huang13}. This work is supported by the NSFC (Grants No. 11271237, No. 61228305, No. 61303009, No. 11201279, and No. 11401361), the Higher School Doctoral Subject Foundation of Ministry of Education of China (Grant No. 20130202110001), and Fundamental Research Funds for the Central Universities (Grants No. GK201502004, No. GK201503017 and No. 2016CBY003).

\appendix
\section{The critical value of $q$ for two-qubit state}
In this section, we will discuss the analytic formula of TEE in two-qubit systems. Let us consider the monotonicity and convexity of $f_q(\mathcal{C}^2)$ as a function of $\mathcal{C}$, where $0\leq\mathcal{C}\leq1.$  First, from Ref.~\cite{Kim10}, we obtain that $f_q(\mathcal{C}^2)$ is a monotonic increasing function of $\mathcal{C}$ for any $q>0$ and $0\leq\mathcal{C}\leq1.$ Second, we will consider the convexity of $f_q(\mathcal{C}^2)$ as a function of $\mathcal{C}$. Kim has proven the convexity of $f_q(\mathcal{C}^2)$ as a function of $\mathcal{C}$ for $1\leq q\leq4$ and the non-convexity of $f_q(\mathcal{C}^2)$ as a function of $\mathcal{C}$ for $q\geq5$~\cite{Kim10}. Thus, we only consider the situation of $0<q<1$ and $4<q<5$, respectively. The function $f_{q}(\mathcal{C}^2)$ is defined as
\begin{equation}\label{}
f_q(\mathcal{C}^2)=\frac{1}{q-1}[1-(\frac{1+\sqrt{1-\mathcal{C}^2}}{2})^q-(\frac{1-\sqrt{1-\mathcal{C}^2
}}{2})^q].
\end{equation}

The second derivative of $f_q(\mathcal{C}^2)$ is
\begin{eqnarray}
\frac{\partial^2f_q(\mathcal{C}^2)}{\partial \mathcal{C}^2}&=&\alpha[\frac{(1+\sqrt{1-\mathcal{C}^2})^{q-1}}{(1-\mathcal{C}^2)^{3/2}} \nonumber\\
&-&\frac{\mathcal{C}^2(q-1)(1+\sqrt{1-\mathcal{C}^2})^{q-2}}{(1-\mathcal{C}^2)}-\frac{(1-\sqrt{1-\mathcal{C}^2})^{q-1}}{(1-\mathcal{C}^2)^{3/2}}\nonumber\\
&-&\frac{\mathcal{C}^2(q-1)(1-\sqrt{1-\mathcal{C}^2})^{q-2}}{(1-\mathcal{C}^2)}]\nonumber
\end{eqnarray}
where $\alpha=\frac{q}{2^q(q-1)}$. For the region $0<q<1$, the convexity of $f_q(\mathcal{C}^2)$ holds if $\frac{\partial^2}{\partial \mathcal{C}^2}f_q(\mathcal{C}^2)\geq 0$ for any concurrence $\mathcal{C}.$ To find the region of $q$, we analyze the condition $\frac{\partial^2}{\partial \mathcal{C}^2}f_q(\mathcal{C}^2)=0$. Numberical calculation shows that the value of $q$ increases monotonically along with the increase of concurrence $\mathcal{C}.$ As showed in FIG.7, there may exist a critical point $q_{c_1}$ corresponds to the limit $\mathcal{C}\to 1$ and the requirement that
\begin{eqnarray}
\lim_{\mathcal{C}\to 1}\frac{\partial^2f_q(\mathcal{C}^2)}{\partial \mathcal{C}^2}=0.
\end{eqnarray}
After some straightforward calculation, we derive the following equality
\begin{eqnarray}
-2(q-1)(q^2-5q+3)=0.
\end{eqnarray}
The critical point of the region $0<q<1$ is $q_{c_1}=\frac{5-\sqrt{13}}{2}\approx 0.697$. The second derivative is nonnegative in this region is $q_{c_1}\leq q<1$.
For the region $4<q<5$, we obtain the critical point $q_{c_2}$ by the similar method. As showed in FIG.8, it shows that the value of $q$ decrease monotonically along with the increase of concurrence $\mathcal{C}$, the critical point $q_{c_2}$ can be obtain by the limit $\lim_{\mathcal{C}\to 1}\frac{\partial^2}{\partial \mathcal{C}^2}f_q(\mathcal{C}^2)=0.$ Thus the critical point of the region $4<q<5$ is $q_{c_2}=\frac{5+\sqrt{13}}{2}\approx 4.302$. The second derivative is nonnegative in this region is $4<q\leq q_{c_1}$.
Therefore, the second derivative is nonnegative for $q_{c_1}\leq q\leq q_{c_2}$ in the region of $0<q<5$. The analytic formula of TEE in two-qubit systems is in this region.

\begin{figure}
\begin{center}
\begin{minipage}[c]{0.25\textwidth}
\centering
\includegraphics[angle=0,width=4cm,height=4cm]{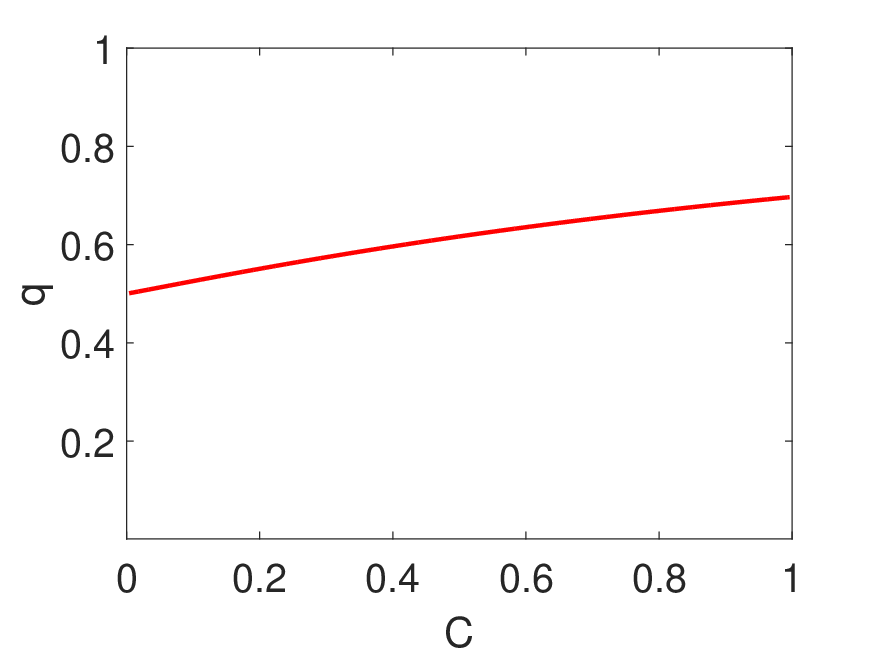}
\centering
\caption{(color online) the condition $\frac{\partial^2}{\partial \mathcal{C}^2}f_q(\mathcal{C}^2)=0$ for $q\in[0,1]$.}
\label{fig:levfig}
\end{minipage}%
\begin{minipage}[c]{0.25\textwidth}  \centering\includegraphics[angle=0,width=4cm,height=4cm]{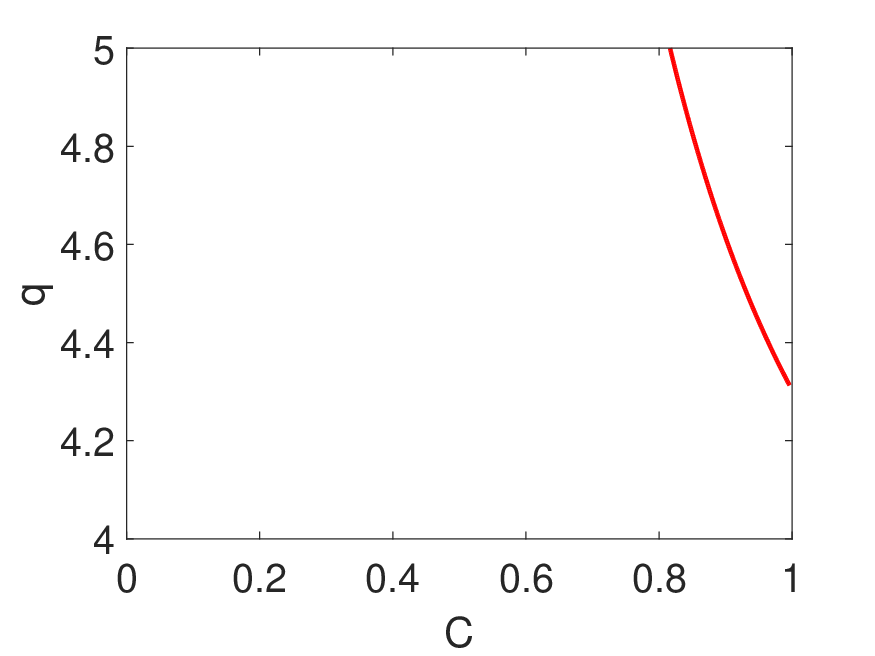}
\caption{(color online) the condition $\frac{\partial^2}{\partial \mathcal{C}^2}f_q(\mathcal{C}^2)=0$ for $q\in[4,5]$. }
\label{fig:levfig}
\end{minipage}
\end{center}
\end{figure}

\section{$f_{q}^2(\mathcal{C}^2)$ is an increasing monotonic and convex function of the squared concurrence $\mathcal{C}^2$}
Firstly, let's consider the monotonicity of the function $f_{q}(x)$, $f_{q}(x)$ is defined as
\begin{equation}\label{}
f_q(x)=\frac{1}{q-1}[1-(\frac{1+\sqrt{1-x}}{2})^q-(\frac{1-\sqrt{1-x}}{2})^q].
\end{equation}
$f_{q}^2(\mathcal{C}^2)$ is an increasing monotonic function of the squared concurrence $\mathcal{C}^2$ is equivalent to the first derivative $\frac{\partial}{\partial x}f_q^2(x)\geq 0$ with $q\in[\frac{5-\sqrt{13}}{2},\frac{5+\sqrt{13}}{2}]$ and $x=\mathcal{C}^2$. After some calculation, we have
\begin{eqnarray}
\frac{\partial f_q^2(x)}{\partial x}=\frac{qf_q(x)}{2^q\sqrt{1-x}}\frac{A^{q-1}-B^{q-1}}{q-1},
\end{eqnarray}
where $A=1+\sqrt{1-x}$ and $B=1-\sqrt{1-x}$. It is easy to check that $\frac{\partial}{\partial x}f_q^2(x)$ is nonnegative for $q\geq 0$. Thus, $f_{q}^2(x)$ is an increasing monotonic function of $x$ for $q\in[\frac{5-\sqrt{13}}{2},\frac{5+\sqrt{13}}{2}]$.

Secondly, the squared Tsallis $q$-entropy entanglement $f_{q}^2(\mathcal{C}^2)$ is a convex function of the squared concurrence $\mathcal{C}^2$ for $q\in[\frac{5-\sqrt{13}}{2},\frac{5+\sqrt{13}}{2}]$, which is equivalent to the second derivative $\frac{\partial^2}{\partial x^2}f_q^2(x)\geq 0.$ Thus, we define the function
\begin{equation}\label{lqx}
l_q(x)=\frac{\partial^2f_q^2(x)}{\partial x^2}
\end{equation}
on the domain $D=\{(x,q)|x\in[0,1], q\in[\frac{5-\sqrt{13}}{2},\frac{5+\sqrt{13}}{2}]\}$. After a straightforward calculation, we have
\begin{eqnarray}
l_q(x)&=&\nonumber\frac{q^2}{8(1-x)}\frac{(A^{q-1}-B^{q-1})^2}{2^{2(q-1)}(q-1)^2}+\frac{f_q(x)}{q-1}\\\nonumber
&\times& [\frac{q(1-q)}{8(1-x)}\frac{A^{q-2}+B^{q-2}}{2^{q-2}}+\frac{q}{4(1-x)^{3/2}} \\\nonumber
&\times& \frac{A^{q-1}-B^{q-1}}{2^{q-1}}].
\end{eqnarray}
The intermediate value theorem tell us if a continuous function on the domain have two values with opposite signs, there must exist a root on the domain. The function $l_q(x)$ is continuous on the domain $D$, and we plot the solution of $l_q(x)=0$. As shown in FIG.9, no point exists on the domain $D$ such that $l_q(x)=0$. Thus the value of $l_q(x)$ on the domain $D$ have the some sign. When $q\to1$, $f_{q}^2(\mathcal{C}^2)$ converges to squared entanglement of formation, which second derivative is positive~\cite{Bai1401}. Therefore, $l_q(x)$ is positive on the domain $D$. We have plot the function $l_q(x)$ on the domain $D$ in FIG.10.

\begin{figure}[htbp]
\begin{center}
\includegraphics[width=6cm,height=6cm]{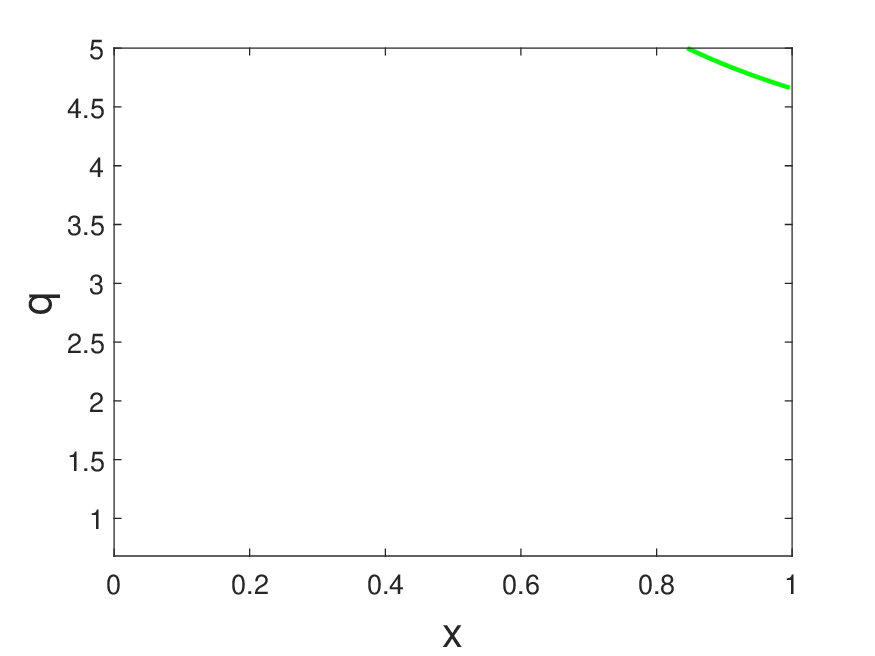}
\caption{(color online) The solution of $l_q(x)=0$ on the domain $D$.
}
\end{center}
\end{figure}

\begin{figure}[htbp]
\begin{center}
\includegraphics[width=6cm,height=6cm]{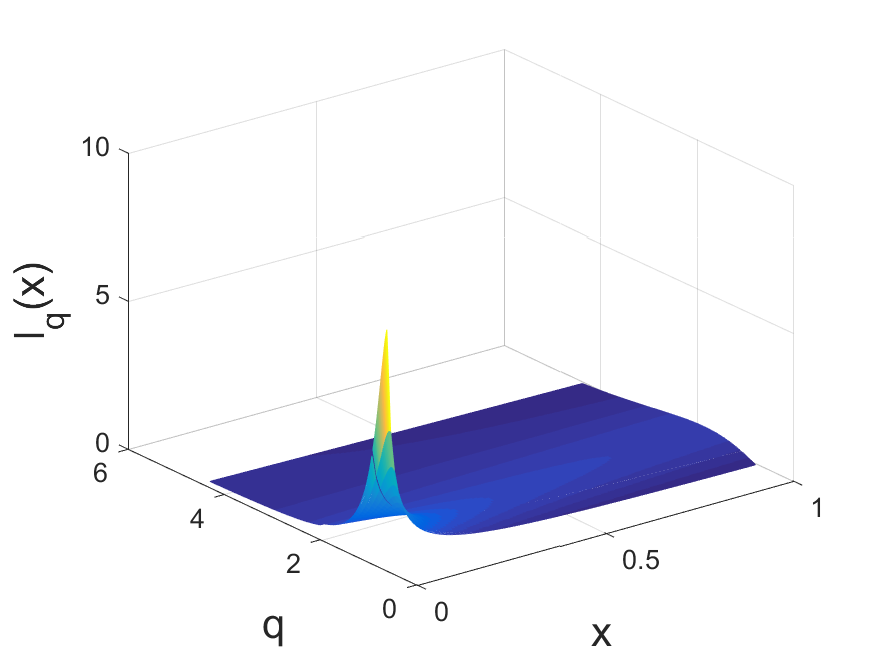}
\caption{(color online) The function $l_q(x)$ is positive on the domain $D$.
}
\end{center}
\end{figure}

\section{$f_{q}(\mathcal{C}^2)$ is an increasing monotonic and concave function of the squared concurrence $\mathcal{C}^2$}
$f_{q}(\mathcal{C}^2)$ is an increasing monotonic function if the first derivative $\frac{\partial}{\partial x}f_q(x)$ is nonnegative.
\begin{eqnarray}
\frac{\partial f_q(x)}{\partial x}=\frac{q}{2^{q+1}\sqrt{1-x}}\frac{A^{q-1}-B^{q-1}}{q-1},
\end{eqnarray}
which is nonnegative for $q\geq \frac{5-\sqrt{13}}{2}$ and $0 \leq x \leq 1$. Namely, $f_{q}(\mathcal{C}^2)$ is an increasing monotonic function of the squared concurrence $\mathcal{C}^2$.

The concavity of function $f_{q}(\mathcal{C}^2)$ is decided by the second derivative $\frac{\partial^2}{\partial x^2}f_q(x)$, and we define the function
\begin{equation}\label{lqx}
g_q(x)=\frac{\partial^2f_q(x)}{\partial x^2}
\end{equation}
on the domain $D=\{(x,q)|x\in[0,1], q\in[\frac{5-\sqrt{13}}{2},\frac{5+\sqrt{13}}{2}]\}$. We have

\begin{eqnarray}
g_q(x)&=&\frac{q}{2^{q+2}(q-1)}[\frac{A^{q-2}}{1-x}(\frac{A}{\sqrt{1-x}}+(1-q))\nonumber\\
&-&\frac{B^{q-2}}{1-x}(\frac{B}{\sqrt{1-x}}-(1-q))].
\end{eqnarray}
In order to find the region of $q$ such that $\frac{\partial^2}{\partial x^2}f_q(x)\leq 0$, we consider equality $\frac{\partial^2}{\partial x^2}f_q(x)=0$ and plot the solution. As showed in FIG.11, the equality holds on the domain only if $q=2, 3$, which cut the domain $D$ into three domains: $D_1=\{(x,q)|x\in[0,1], q\in[\frac{5-\sqrt{13}}{2},2]\}$, $D_2=\{(x,q)|x\in[0,1], q\in(2,3]\}$ and $D_3=\{(x,q)|x\in[0,1], q\in(3,\frac{5+\sqrt{13}}{2}]\}$. The corresponding functions for $q=2, 3$ are
\begin{eqnarray}
f_2(x)=\frac{x}{2},\qquad f_3(x)=\frac{3x}{8},
\end{eqnarray}
where $0\leq x\leq 1.$  The intermediate value theorem tell us if a continuous function have two values on the domain with opposite signs, there must exist a root on the domain. The function $\frac{\partial^2}{\partial x^2}f_q(x)$ is a continuously function on the domain $D=D_1\cup D_2\cup D_3$. Therefore, we can consider the condition of $q=1$, $q=\frac{5}{2}$ and $q=4$ which on the domain $D_1$, $D_2$ and $D_3$ respectively.
When $q=1$, the TEE converges to entanglement of formation, it have been proven in Re.~\cite{Bai1408} that $g_1(x)< 0$ for $x\in[0,1]$. Thus, $g_q(x)< 0$ is nonpositive on the domain $D_1$ and equality holds only if $q=2$. When $q=\frac{5}{2}$, we have

\begin{eqnarray}
g_\frac{5}{2}(x)=-\frac{15}{64\sqrt2}\frac{A^\frac{1}{2}+B^\frac{1}{2}}{1-x}+\frac{5}{32\sqrt2}\frac{A^\frac{3}{2}-B^\frac{3}{2}}{(1-x)^\frac{3}{2}}.
\end{eqnarray}
It's easy to check that $\lim_{x\to0}g_\frac{5}{2}(x)=\frac{15}{128}>0$ and $\lim_{x\to1}g_\frac{5}{2}(x)=\frac{15}{256\sqrt2}>0$. Thanks to the continuously of $g_\frac{5}{2}(x)$ and the intermediate value theorem, we can obtain that $g_\frac{5}{2}(x)> 0$ for $x\in[0,1]$. Thus, $g_q(x)$ is nonnegativity on the domain $D_2$ and equality holds only if $q=3$. As showed in Fig. 12, the function $g_q(x)$ is nonnegativity on the domain $D_2$. When $q=4$, we have

\begin{eqnarray}
f_4(x)=\frac{8x-x^2}{24},
\end{eqnarray}
and $g_4(x)=-\frac{1}{12}<0$ for $x\in[0,1]$. Thus, $g_q(x)< 0$ is negativity on the domain $D_3$.
Therefore, the function $f_q(x)$ is concave on the domain $D'=\{(x,q)|x\in[0,1], q\in[\frac{5-\sqrt{13}}{2},2]\cup[3,\frac{5+\sqrt{13}}{2}]
\}$.

\begin{figure}[htbp]
\begin{center}
\includegraphics[width=6cm,height=6cm]{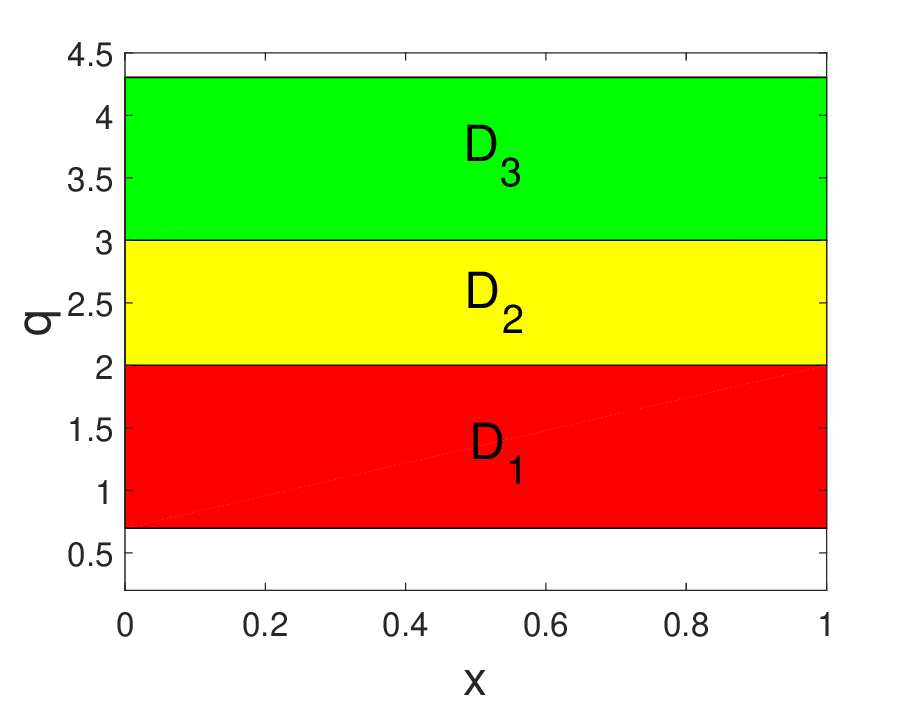}
\caption{(color online) the condition $g_q(x)=0$, which holds on the domain only if $q=2, 3$ and cut the domain $D$ into three domains: $D_1$ (red color), $D_2$ (yellow color) and $D_3$ (green color).}
\end{center}
\end{figure}

\begin{figure}[htbp]
\begin{center}
\includegraphics[width=6cm,height=6cm]{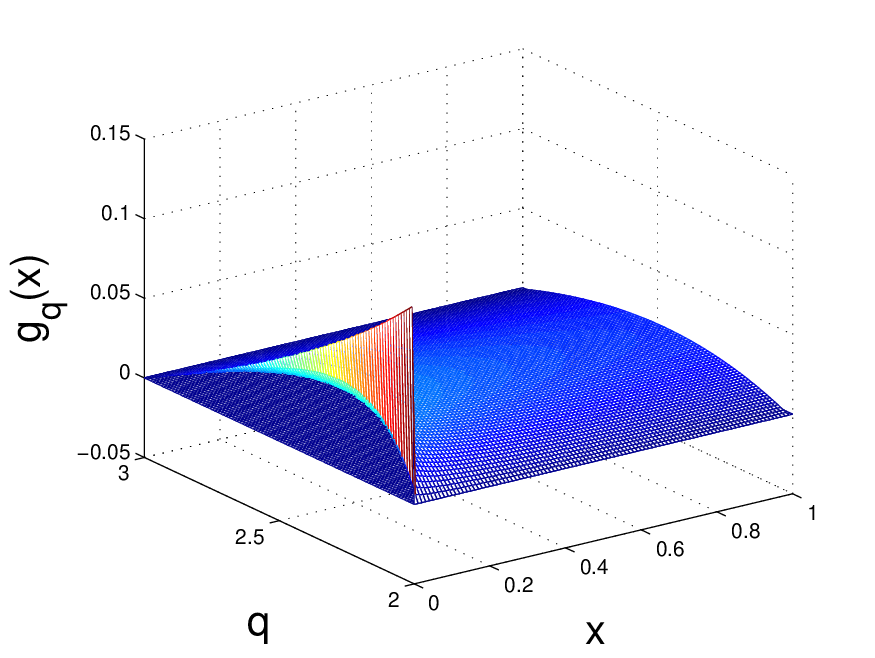}
\caption{(color online) $g_q(x)$ is nonnegativity on the domain $D_2$.}
\end{center}
\end{figure}

\section{Monogamy of the $\alpha$th power of TEE }
Assuming $\sum_{i=2}^{N-1}\mathcal{T}_{q}^2(\rho_{A_1A_i})\geq\mathcal{T}_{q}^2(\rho_{A_1A_N})$, from the Eq.~(\ref{eq:The1}) we have

\begin{eqnarray}\label{}
\mathcal{T}^{\alpha}_{q}(\rho_{A_1|A_2\ldots A_N})&\geq&\nonumber (\mathcal{T}^{2}_{q}(\rho_{A_1A_2})+\cdots+\mathcal{T}^{2}_{q}(\rho_{A_1A_N}))^{\frac{\alpha}{2}}\\\nonumber
&=& (\sum_{i=2}^{N-1}\mathcal{T}_{q}^2(\rho_{A_1A_i}))^{\frac{\alpha}{2}}(1+\frac{\mathcal{T}^{2}_{q}(\rho_{A_1A_N})}{\sum_{i=2}^{N-1}\mathcal{T}_{q}^2(\rho_{A_1A_i})})^{\frac{\alpha}{2}} \\\nonumber
&\geq& (\sum_{i=2}^{N-1}\mathcal{T}_{q}^2(\rho_{A_1A_i}))^{\frac{\alpha}{2}}(1+(\frac{\mathcal{T}^{2}_{q}(\rho_{A_1A_N})}{\sum_{i=2}^{N-1}\mathcal{T}_{q}^2(\rho_{A_1A_i})})^{\frac{\alpha}{2}}) \\\nonumber
&=& (\sum_{i=2}^{N-1}\mathcal{T}_{q}^2(\rho_{A_1A_i}))^{\frac{\alpha}{2}}+\mathcal{T}^{\alpha}_{q}(\rho_{A_1A_N}) \\\nonumber
&\geq& \mathcal{T}^{\alpha}_{q}(\rho_{A_1A_2})+\cdots+\mathcal{T}^{\alpha}_{q}(\rho_{A_1A_N}),
\end{eqnarray}
where the second inequality holds is due to the property $(1+x)^t\geq 1+x^t$, where $0\leq x\leq1$ and $t\geq1$, the third inequality holds is due to the property $(\sum x_i^2)^{\frac{\alpha}{2}}\geq\sum x_i^{\alpha}$, where $0\leq x_i\leq1$ and $\alpha\geq2.$

\end{document}